\documentclass[12pt,prd,tightenlines,nofootinbib,showpacs,showkeys,setspace]{revtex4}
\newcommand{\be}{\begin{equation}}
\newcommand{\ee}{\end{equation}}
\newcommand{\bdis}{\begin{displaymath}}
\newcommand{\edis}{\end{displaymath}}
\newcommand{\bga}{\begin{equation}\begin{gathered}}
\newcommand{\ega}{\end{gathered}\end{equation}}
\newcommand{\mathsym}[1]{{}}
\newcommand{\unicode}[1]{{}}
\usepackage{bm}
\usepackage{graphics}
\usepackage{rotating}
\usepackage{epsfig}
\usepackage{amsmath}
\usepackage{amsfonts}
\usepackage{setspace}
\usepackage{amssymb}

\begin{document}
\title{
Exclusive double $B_c$ meson production
from $e^+e^-$ annihilation into two virtual photons}
\author{\firstname{A.V.} \surname{Berezhnoy}}
\affiliation{SINP MSU, Vorob'evy Gory, 119992, Moscow, Russia}
\author{\firstname{A.P.} \surname{Martynenko}}
\author{\firstname{F.A.} \surname{Martynenko}}
\author{\firstname{O.S.} \surname{Sukhorukova}}
\affiliation{Samara National Research University, Moskovskoye Shosse 34, 443086, Samara, Russia}

\begin{abstract}
We calculate cross sections of a pair $B_c$ meson production on the basis of two-photon 
mechanism from electron-positron annihilation. We investigate the production cross
sections in nonrelativistic approximation and with the account of relativistic corrections.
Relativistic production amplitudes of S-wave pair pseudoscalar, vector and pseudoscalar+vector
$B_c$-mesons are constructed on the basis of relativistic quark model. 
Numerical values of the production cross sections are obtained at different 
center-of-mass energies. The comparison of one-photon and two-photon annihilation contributions
is presented.
\end{abstract}



\maketitle

\section{Introduction}

The production of bound states of heavy quarks (double-heavy mesons and baryons) in such fundamental 
processes as electron-positron annihilation and proton-proton interaction is of great importance for testing the perturbative and non-perturbative aspects of quantum chromodynamics and the theory of bound states. 
Among these reactions stands out the process of exclusive pair production of heavy quarkonia and double heavy 
baryons, since a pair of bound states of heavy quarks is produced in it and the binding effects of quarks 
are significantly enhanced. The activity in the theoretical study of the pair charmonium production 
in electron-positron annihilation was largely connected with the experimental results obtained by the Belle and BaBar collaborations \cite{belle,babar,pahlova,brambilla2011}, which differed significantly from the predictions of non-relativistic quantum chromodynamics \cite{bl1,qiao,chao1}. 
During a short period of research, it has been shown that the theoretical results for the pair production 
of charmonium can be reconciled with experimental data, taking into account perturbative corrections 
of order $O(\alpha_s)$ to the production amplitudes and relativistic corrections, due to the relative 
motion of heavy quarks \cite{bodwin1,bodwin2,bodwin3,chao2,chao3,bll1,bll2,em2006,ji,jia,gong,akl1,efgm2009}. 
Along with the mechanism of one-photon annihilation, the production process of a pair 
of charmoniums in two-photon annihilation was also studied at this time \cite{bbl1}. Despite the fact that such 
contributions contain an additional small factor $\alpha^2/\alpha_s^2$, nevertheless, 
the fragmentation contribution has a structure that compensates for the suppression 
of the production amplitude by coupling constants.

In our work \cite{apm2016}, we used the developed methods of studying the reactions of pair
exclusive quarkonia production in the case of $B_c$-mesons. Our results \cite{apm2016} show
that, as in the case of $(c\bar b)$ mesons, relativistic effects substantially change
the magnitude of the cross sections for the production of a pair of $B_c$-mesons. 
In \cite{akl2}, the calculation was performed for 
one-loop corrections in the pair-production of $B_c$-mesons in a one-photon annihilation. 
An approximation was used in which the relative motion of heavy quarks was not taken into account. 
In this paper, we explore the contribution of
two-photon annihilation process in reactions of the production of a pair of $B_c$-mesons.
We calculate the cross sections for the production of a pair of $B_c$-mesons in the nonrelativistic 
approximation and with the account of relativistic corrections. A comparison of numerical contributions
of one-photon and two-photon mechanisms in total cross section is performed.

While extensive literature is devoted to the issues of the production of single $B_c$ mesons \cite{gklt}, 
the problem of the production of a pair of $B_c$ mesons or heavy diquarks $(bc)$ in different reactions 
has been discussed to a much lesser extent \cite{apm2016,mt,mt1,mt2}. 
This is mainly due to the lack of a clear experimental perspective to observe 
a sufficient number of such events in existing experiments.
This work continues our research of exclusive double heavy meson production in $e^+e^-$ annihilation.
Our approach to calculating the observed cross sections for the production of a pair of mesons
is based on methods of relativistic quark model (RQM) and perturbative quantum chromodynamics
\cite{em2006,efgm2009,apm2016,em2010,apm2005,apm2007}. In this approach we can take into account 
relativistic effects in the construction of relativistic production amplitudes, 
relativistic production cross sections,
and in the description of bound states of heavy quarks through the use of the corresponding 
quark interaction operator. We can say that in this approach we have
microscopic picture of the photon, quark and gluon interaction at different stages of meson production.
The approach based on relativistic quark model allows you to perform a self-consistent calculation 
of various theoretical parameters, which ultimately determine the total numerical values of the production
cross sections.

\section{General formalism}

\begin{figure}[htbp]
\centering
\includegraphics[scale=1.0]{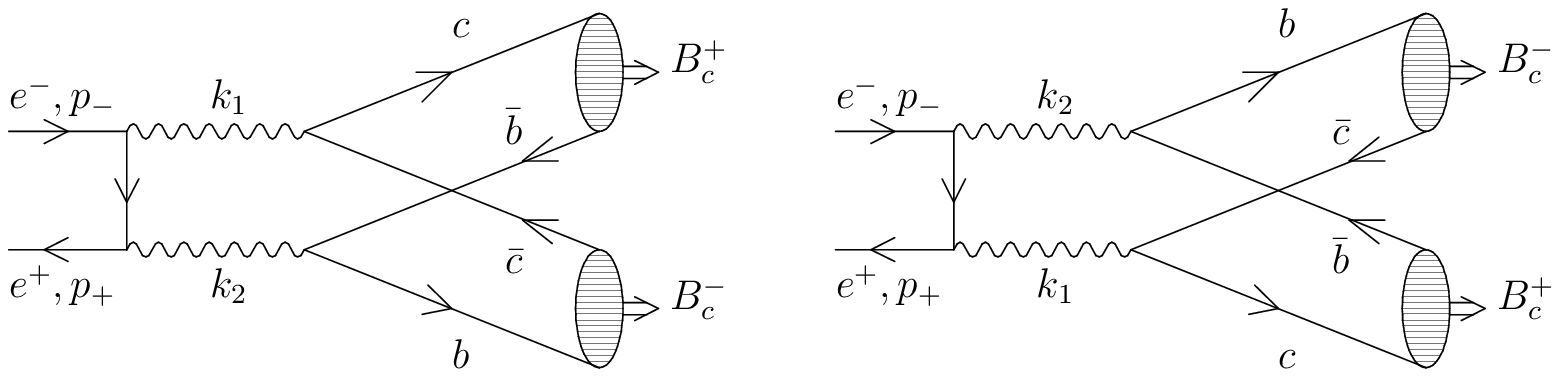}
\caption{The pair $B_c$-meson production amplitudes in $e^+e^-$ annihilation. $B^+_{c}$ and $B^{-}_{c}$
denote the $B_c$-meson states with spin 0. Wavy lines show
the virtual photons with four momenta $k_1$ and $k_2$.}
\label{fig:fig1}
\end{figure}

Two-photon annihilation amplitudes leading to the production of a pair of $B_c$ mesons,
at leading order are presented in Fig.~\ref{fig:fig1}. 
The production process can be divided into two stages. At the first stage, a pair of virtual photons is formed.
At the second stage, each virtual photon produces a pair of quark-antiquark, which then with some probability 
combine into $B_c$ mesons with definite spin.
To properly take into account the quark binding effects and relativistic corrections, 
we express quark four-momenta in terms of the total and relative four-momenta in the form:
\begin{equation}
\label{eq:pq}
p_1=\eta_{1}P+p,~p_2=\eta_{2}P-p,~(p\cdot P)=0,~\eta_{1,2}=\frac{M_{B_{\bar bc}}^2\pm m_1^2\mp m_2^2}
{2M_{B_{\bar bc}}^2},
\end{equation}
\begin{displaymath}
q_1=\rho_{1}Q+q,~q_2=\rho_{2}Q-q,~(q\cdot Q)=0,~\rho_{1,2}=\frac{M_{B_{b\bar c}}^2\pm m_1^2\mp m_2^2}
{2M_{B_{b\bar c}}^2},
\end{displaymath}
where $M_{B_{\bar bc}}$ is the mass of pseudoscalar or vector $B_c^+$ ($B_c^{\ast +}$) meson consisting 
of $\bar b$-antiquark and $c$-quark.
$P(Q)$ are the total four-momenta of mesons $B_c^+$ and $B_c^{\ast -}$, relative quark four-momenta $p=L_P(0,{\bf p})$ and
$q=L_P(0,{\bf q})$ are obtained from the rest frame four-momenta $(0,{\bf p})$ and $(0,{\bf q})$ by the
Lorentz transformation to the system moving with the momenta $P$ and $Q$.
It can be noted that a good choice of coefficients in the formulas \eqref{eq:pq} leads to the orthogonality 
of the total and relative four-momenta. In turn, the virtual photon momenta $k_{1,2}$ can also be expressed through
P, Q, p, q in the form:
\begin{equation}
\label{eq:k1k2}
k_1^2=(p_1+q_1)^2=(\eta_1 P+\rho_1 Q +p+q)^2,~~~k_2^2=(p_2+q_2)^2=(\eta_2 P+\rho_2 Q -p-q)^2,
\end{equation}
and the virtuality of each photon is large.
We note that in our approach, quarks are not in an intermediate state on the mass shell, since after 
their production there is always an interaction between them, and there is a symmetric escape
of particles beyond the mass shell:
$p_{1,2}^2=\eta_{1,2}^2P^2-{\bf p}^2=\eta_{1,2}^2M_{B_{\bar bc}}^2-{\bf p}^2\not= m_{1,2}^2$,
$p_1^2-m_1^2=p_2^2-m_2^2$. Since we are discussing the creation of a pair of S-wave states, 
in the case of a pair of pseudoscalar or pair of vector mesons $\eta_{1,2}=\rho_{1,2}$,
and at the production of pseudoscalar and vector mesons $\eta_{1,2}\approx \rho_{1,2}$ with
good accuracy. Indeed, the numerical values of these coefficients can be obtained by choosing 
the masses of the mesons $M_{\cal P}=6.2749$ GeV \cite{pdg}, $M_{\cal V}=6.332$ GeV \cite{efg2003}: 
$\eta_1=0.228$, $\rho_1=0.233$, $\eta_2=0.772$, $\rho_2=0.767$ (hereinafter, the indices denote 
pseudoscalar ${\cal P}$ and vector ${\cal V}$ states.).

The construction of the relativistic amplitude of the two-photon pair production includes 
a number of intermediate steps.
In the color singlet model the production amplitude of four quarks and anti-quarks 
presented in first  Feynman diagram takes the form:
\begin{equation}
\label{eq:t1}
{\cal T}_1(p,q,P,Q)=\frac{16\pi^2\alpha^2 Q_cQ_b}{k_1^2k_2^2}\bar v(p_+)\frac{\gamma^\alpha
(\hat p_--\hat k_1+m_e)\gamma^\beta}
{(p_--k_1)^2-m_e^2} u(p_-)\bar u^i_1(p_1)\gamma^\beta v^i_1(q_1)
\bar u^j_2(q_2)\gamma^\alpha v^j_2(p_2),
\end{equation}
where $p_-$, $p_+$ are four momenta of electron and positron,
$u^i_{1,2}$, $v^j_{1,2}$ are wave functions of quarks and anti-quarks playing the role of projection
operators on positive energy states.
Accounting for color part of the meson wave function ($\delta_{ij}/\sqrt{3}$) we obtain total color factor
in the amplitude \eqref{eq:t1} equal to 1. In quasipotential approach we can express the amplitude of
the reaction $e^++e^-\to B^+_{\bar bc}(B^{\ast +}_{\bar bc})+B^-_{\bar c b}(B^{\ast -}_{\bar bc})$ 
as a convolution of ${\cal T}_1(p,q,P,Q)$ and quasipotential wave functions of produced quark bound states:
\begin{equation}
\label{eq:amp1}
{\cal M}_1(p_-,p_+,P,Q)=\int\frac{d{\bf p}}{(2\pi)^3}\bar\Psi_{\cal P}(p,P)
\int\frac{d{\bf q}}{(2\pi)^3}\bar\Psi_{\cal V}(q,Q){\cal T}_1(p,q,P,Q).
\end{equation}
In this matrix element we have the wave functions of mesons moving with four-momenta P and Q.
The transformation law of the bound state wave function from the rest frame $\Psi_0({\bf p})$ to the
moving one with four-momentum P was derived in the Bethe-Salpeter approach in \cite{brodsky}
and in quasipotential method in \cite{faustov}. We use the quasipotential method, so
the wave function transformation looks as follows:
\begin{equation}
\label{eq:tran1}
\Psi_{P}^{\rho\omega}({\bf p})=D_1^{1/2,~\rho\alpha}(R^W_{L_{P}})
D_2^{1/2,~\omega\beta}(R^W_{L_{P}})\Psi_{0}^{\alpha\beta}({\bf p}),
\end{equation}
\begin{displaymath}
\bar\Psi_{P}^{\lambda\sigma}({\bf p})
=\bar\Psi^{\varepsilon\tau}_{0}({\bf p})D_1^{+~1/2,~\varepsilon
\lambda}(R^W_{L_{P}})D_2^{+~1/2,~\tau\sigma}(R^W_{L_{P}}),
\end{displaymath}
where $R^W$ is the Wigner rotation, $L_{P}$ is the Lorentz boost
from the meson rest frame to a moving one, and the rotation matrix $D^{1/2}(R)$ is defined by
\begin{equation}
\label{eq:tran2}
{1 \ \ \,0\choose 0 \ \ \,1}D^{1/2}_{1,2}(R^W_{L_{P}})=
S^{-1}({\bf p}_{1,2})S({\bf P})S({\bf p}),
\end{equation}
where the explicit form for the Lorentz transformation matrix of the four-spinor is
\begin{equation}
\label{eq:tran3}
S({\bf p})=\sqrt{\frac{\epsilon(p)+m}{2m}}\left(1+\frac{(\bm{\alpha}
{\bf p})} {\epsilon(p)+m}\right).
\end{equation}
Further transformations of the amplitude \eqref{eq:amp1} can be carried out by means of the following relations:
\begin{equation}
\label{eq:tran4}
S_{\alpha\beta}(\Lambda)u^\lambda_\beta(p)=\sum_{\sigma=\pm 1/2}
u^{\sigma}_\alpha(\Lambda p)D^{1/2}_{\sigma\lambda}(R^W_{\Lambda p}),
\end{equation}
\begin{displaymath}
\bar u^\lambda_\beta(p)S^{-1}_{\beta\alpha}(\Lambda)=\sum_{\sigma=\pm 1/2}
D^{+~1/2}_{\lambda\sigma}(R^W_{\Lambda p})\bar u^\sigma_\alpha(\Lambda p).
\end{displaymath}
Using also the transformation property of the Dirac bispinors to the rest frame
\begin{eqnarray}
\label{eq:tran5}
\bar u_1({\bf p})=\bar u_1(0)\frac{(\hat
p'_1+m_1)}{\sqrt{2\epsilon_1({\bf p}) (\epsilon_1({\bf
p})+m_1)}},~~p'_1=(\epsilon_1,{\bf p}),\cr\cr v_2(-{\bf
p})=\frac{(\hat p'_2-m_2)}{\sqrt{2\epsilon_2({\bf
p})(\epsilon_2({\bf p})+ m_2)}}v_2(0),~~p'_2=(\epsilon_2,-{\bf p}),
\end{eqnarray}
we can introduce the projection operators $\hat\Pi^{{\cal P},{\cal V}}$
onto the states $(c\bar b)$, $(b\bar c)$ with total spin 0 and 1 as follows:
\begin{equation}
\label{eq:tran6}
\hat\Pi^{{\cal P},{\cal V}}=[v_2(0)\bar
u_1(0)]_{S=0,1}=\gamma_5(\hat\epsilon^\ast)\frac{1+\gamma^0}{2\sqrt{2}}.
\end{equation}

As a result of such transformations the total amplitude of pair $B_c$ meson production
can be written in the form:
\begin{equation}
\label{eq:amp2}
{\cal M}(p_-,p_+,P,Q)=\int\frac{d{\bf p}}{(2\pi)^3}\int\frac{d{\bf q}}{(2\pi)^3}
\frac{4\pi^2\alpha^2Q_cQ_b}{k_1^2k_2^2}\sqrt{M_{B_{\bar bc}}M_{B_{b\bar c}}}
\bar v(p_+)\Bigl[\frac{\gamma^\alpha(\hat p_--\hat k_1+m_e)\gamma^\beta}{(p_--k_1)^2-m_e^2}+
\end{equation}
\begin{displaymath}
\frac{\gamma^\beta(\hat p_--\hat k_2+m_e)\gamma^\alpha}{(p_--k_2)^2-m_e^2}\Bigr]u(p_-)
Sp\left\{\Psi^{\cal P}_{B_{\bar bc}}(p,P)\gamma^{\beta}\Psi^{\cal V}_{B_{b\bar c}}(q,Q)\gamma^\alpha\right\},
\end{displaymath}
where $s$ is the center-of-mass energy,
a superscript ${\cal P}$ indicates a pseudoscalar $B_c$ meson, a superscript ${\cal V}$ indicates a vector
$B_c$ meson, $\alpha$ is the fine structure constant.
The transition wave functions $\Psi^{\cal V}_{B_{b\bar c}}(q,Q)$ and $\Psi^{\cal P}_{B_{\bar bc}}(p,P)$
(form factors) have the following form:
\begin{eqnarray}
\label{eq:amp3}
\Psi^{\cal P}_{B_{\bar bc}}(p,P)&=&\frac{\Psi^0_{B_{\bar bc}}({\bf p})}{
\sqrt{\frac{\epsilon_1(p)}{m_1}\frac{(\epsilon_1(p)+m_1)}{2m_1}
\frac{\epsilon_2(p)}{m_2}\frac{(\epsilon_2(p)+m_2)}{2m_2}}}
\left[\frac{\hat v_1-1}{2}+\hat
v_1\frac{{\bf p}^2}{2m_2(\epsilon_2(p)+ m_2)}-\frac{\hat{p}}{2m_2}\right]\cr
&&\times\gamma_5(1+\hat v_1) \left[\frac{\hat
v_1+1}{2}+\hat v_1\frac{{\bf p}^2}{2m_1(\epsilon_1(p)+
m_1)}+\frac{\hat{p}}{2m_1}\right],
\end{eqnarray}
\begin{eqnarray}
\label{eq:amp4}
\Psi^{\cal V}_{B^\ast_{b\bar c}}(q,Q)&=&\frac{\Psi^0_{B^\ast_{b\bar c}}({\bf q})}
{\sqrt{\frac{\epsilon_1(q)}{m_1}\frac{(\epsilon_1(q)+m_1)}{2m_1}
\frac{\epsilon_2(q)}{m_2}\frac{(\epsilon_2(q)+m_2)}{2m_2}}}
\left[\frac{\hat v_2-1}{2}+\hat v_2\frac{{\bf q}^2}{2m_1(\epsilon_1(q)+
m_1)}+\frac{\hat{q}}{2m_1}\right]\cr &&\times\hat{\varepsilon}_{\cal
V}(Q,S_z)(1+\hat v_2) \left[\frac{\hat v_2+1}{2}+\hat
v_2\frac{{\bf q}^2}{2m_2(\epsilon_2(q)+ m_2)}-\frac{\hat{q}}{2m_2}\right],
\end{eqnarray}
where the symbol hat denotes convolution of four-vector with the Dirac gamma matrices,
$v_1=P/M_{B_{\bar bc}}$, $v_2=Q/M_{B_{b\bar c}}$;
$\varepsilon_{\cal V}(Q,S_z)$ is the polarization vector of the $B^{\ast-}_c(1^-)$ meson,
relativistic quark energies $\epsilon_{1,2}(p)=\sqrt{p^2+m_{1,2}^2}$ and $m_{1,2}$
are the masses of $c$ and $b$ quarks. 
In \eqref{eq:amp2} we have complicated factor including the bound state wave function in the rest frame.
Therefore instead of the substitutions $M_{B_{\bar bc}}=\epsilon_1({\bf p})+\epsilon_2({\bf p})$ and
$M_{B^\ast_{b\bar c}}=\epsilon_1({\bf q})+\epsilon_2({\bf q})$ in the production amplitude we carry out the
integration over the quark relative momenta ${\bf p}$ and ${\bf q}$.
Relativistic wave functions in~\eqref{eq:amp3} and \eqref{eq:amp4} are equal to the product 
of wave functions in the rest frame
$\Psi^0_{B_{\bar bc}}({\bf p})$ and spin projection operators that are
accurate at all orders in $|{\bf p}|/m$. An expression of spin projector in different
form for $(c\bar c)$ system was obtained in \cite{bodwin2002,bodwin2004} where spin projectors are
written in terms of heavy quark momenta $p_{1,2}$ lying on the mass shell.
We can consider \eqref{eq:amp3}-\eqref{eq:amp4} as a transition form factors for
heavy quark-antiquark pair from free state to bound state. 

Relative momenta of heavy quarks $p$ and $q$ enter in the propagators of photons and electron in
intermediate state as well as in transition wave functions~\eqref{eq:amp3} and \eqref{eq:amp4}.
The ratios of relative momenta to s or quark masses are small so we can use an expansion of all quantities
depending on p and q. In the decomposition of the corresponding factors, we take into account 
the terms of the second order in p and q in the form:
\begin{equation}
\label{eq:pr1}
\frac{1}{k_{1,2}^2}=\frac{1}{\eta_{1,2}^2s^2}\left[1\mp \frac{2(pQ+qP)}{\eta_{1,2}s^2}-\frac{(p+q)^2}
{\eta_{1,2}^2s^2}+\cdots\right],
\end{equation}
\begin{equation}
\label{eq:pr2}
\frac{1}{(k_{1,2}-p_-)^2-m_e^2}=-\frac{1}{\eta_1\eta_2s^2}\left[1\pm \frac{2(pQ+qP)}{\eta_{2,1}s^2}\mp \frac{2p_-(p+q)}
{\eta_1\eta_2 s^2}+\frac{(p+q)^2}{\eta_1\eta_2 s^2}+\cdots\right].
\end{equation}
The coefficients $\eta_{1,2}$ include also bound state effects which we express in terms of bound state energies
setting $M_{\cal P}=m_1+m_2+B_{\cal P}$, $M_{\cal V}=m_1+m_2+B_{\cal V}$.
Despite expansions \eqref{eq:pr1}-\eqref{eq:pr2} with respect to relative momenta, the integrals 
in \eqref{eq:amp2} remain convergent 
when taking corrections of second-order in p or q and can be calculated if the $\Psi^0_{B^\ast_{b\bar c}}({\bf p})$
functions are known. 
It is convenient to perform further transformations using the Form package \cite{form} immediately 
when calculating the cross section for the production of a pair.

When calculating the differential cross section, we introduce the angle
$\theta$ between the electron momentum ${\bf p}_e$ and momentum ${\bf P}$ of $B_c$ meson. 
After all the simplifications and the explicit separation of relativistic corrections of the second 
order and the bound state effects, we obtain the differential cross section $d\sigma/d\cos\theta$ ($z=\cos\theta$)
as a function of center-of-mass energy $s$ with a number of parameters containing quark masses, 
binding energies, and relativistic corrections.
The differential cross sections for the production of pair pseudoscalar mesons, pair vector mesons and
pair pseudoscalar plus vector mesons can be written in the following form:
\begin{equation}
\label{eq:sech1}
\frac{d\sigma_{B_{\bar bc}B_{b\bar c}}}{dz}=
\frac{8\pi^3\alpha^4 q_c^2q_b^2M_{B_{\bar bc}}M_{B_{b\bar c}}|{\bf P}|M^4}{r_1^6r_2^6s^{15}}
|\Psi^0_{B_{\bar bc}}(0)|^2|\Psi_{B_{b\bar c}}(0)|^2
\Bigl[
f^{LO}_{B_{\bar bc}B_{b\bar c}}(z,s)+
\end{equation}
\begin{displaymath}
\frac{(B_{B_{\bar bc} }+B_{ B_{b\bar c}})}{2M}f^{bind}_{B_{\bar bc}B_{b\bar c}}(z,s)+
\omega_{\frac{1}{2}\frac{1}{2}}f^{1,rel}_{B_{\bar bc}B_{b\bar c}}(z,s)+
\omega_{01}f^{2,rel}_{B_{\bar bc}B_{b\bar c}}(z,s)+\omega_{10}f^{3,rel}_{B_{\bar bc}B_{b\bar c}}(z,s)
\Bigr],
\end{displaymath}
where $B_{B_{\bar bc}}$ is the binding energy,
$|{\bf P}|=\sqrt{[s^2-(M_{B_{\bar bc}}+M_{B_{b\bar c}})^2][s^2-
(M_{B_{\bar bc}}-M_{B_{b\bar c}})^2]/4s^2}$ is the meson three momentum in center-of-mass frame,
$r_{1.2}=m_{1,2}/(m_1+m_2)=m_{1,2}/M$. The value of bound state wave function at the origin is equal
\begin{equation}
\label{eq:sech2}
\Psi^0_{B_{\bar bc}}(0)=\int \sqrt{\frac{(\epsilon_1(p)+m_1)(\epsilon_2(p)+m_2)}
{2\epsilon_1(p)\cdot 2\epsilon_2(p)}}\Psi^0_{B_{\bar bc}}({\bf p})\frac{d{\bf p}}{(2\pi)^3}.
\end{equation}
In the integral function \eqref{eq:sech2} , we have identified the relativistic factor that usually 
arises in the relativistic quark model. Explicit analytical expressions for the functions 
$f^{1,rel}_{B_{\bar bc}B_{b\bar c}}(z,s)$, $f^{LO}_{B_{\bar bc}B_{b\bar c}}(z,s)$, 
$f^{bind}_{B_{\bar bc}B_{b\bar c}}(z,s)$ are presented in Appendix A.
At the first stage of transformations of the production cross sections in the Forms package, 
we decompose the integrand function in powers of relativistic factors
$C_{nk}=[(m_1-\epsilon_1(p))/(m_1+\epsilon_1(p))]^n
[(m_2-\epsilon_2(q))/(m_2+\epsilon_2(q))]^k$, where $n$, $k$ are integers and half-integers with $n+k\leq 1$.
To preserve the symmetry of the expression on the quark masses
we make following substitution in some expansion
terms: ${\bf p}^2/4m_1m_2\approx\sqrt{(\epsilon_1-m_1)(\epsilon_2-m_2)/(\epsilon_1+m_1)(\epsilon_2+m_2)}
[1+(\epsilon_1-m_1)/(\epsilon_1+m_1)+(\epsilon_2-m_2)/(\epsilon_2+m_2)+...]$.
On the second stage after simplifications based on the symmetry properties of integral
functions when replacing ${\bf p}\to{\bf q}$ we extract in the cross sections specific 
relativistic parameters $\omega_{nk}$.
These parameters can be expressed in terms of momentum integrals $I^{nk}$ and
calculated in the quark model:
\begin{equation}
\label{eq:intnk}
I^{nk}_{{B_{\bar bc},B_{b\bar c}}}=\int_0^\infty q^2R_{{B_{\bar bc},B_{b\bar c}}}(q)\sqrt{\frac{(\epsilon_1(q)+m_1)(\epsilon_2(q)+m_2)}
{2\epsilon_1(q)\cdot 2\epsilon_2(q)}}
\left(\frac{\epsilon_1(q)-m_1}{\epsilon_1(q)+m_1}\right)^n
\left(\frac{\epsilon_2(q)-m_2}{\epsilon_2(q)+m_2}\right)^k dq,
\end{equation}
\begin{equation}
\label{eq:parameter}
\omega^{B_{\bar bc},B_{b\bar c}}_{10}=\frac{I_{B_{\bar bc},B_{b\bar c}}^{10}}{I_{B_{\bar bc},B_{b\bar c}}^{00}},~~~
\omega^{B_{\bar bc},B_{b\bar c}}_{01}=\frac{I_{B_{\bar bc},B_{b\bar c}}^{01}}{I_{B_{\bar bc},B_{b\bar c}}^{00}},~~~
\omega^{B_{\bar bc},B_{b\bar c}}_{\frac{1}{2}\frac{1}{2}}=\frac{I_{B_{\bar bc},B_{b\bar c}}^{\frac{1}{2}\frac{1}{2}}}
{I_{B_{\bar bc},B_{b\bar c}}^{00}},
\end{equation}
where $R_{B_{\bar bc},B_{b\bar c}}(q)$ is the radial wave function of the mesons $B_{\bar bc}$,$ B_{b\bar c}$ in
momentum space. The expansion in \eqref{eq:sech1} can extend and take into account terms of higher order in
p and q. In the process of obtaining the necessary integrand
functions we use different substitutions for ${\bf p}^2$ and ${\bf q}^2$. All of them can be obtained using 
the following expansion
\begin{equation}
\label{eq:expan}
|{\bf p}|=2m_1\left[\sqrt{\frac{\epsilon_1-m_1}{\epsilon_1+m_1}}+\left(\frac{\epsilon_1-m_1}{\epsilon_1+m_1}
\right)^{3/2}+\left(\frac{\epsilon_1-m_1}{\epsilon_1+m_1}\right)^{5/2}+...\right]=
\end{equation}
\begin{equation}
\label{eq:expan1}
2m_2\left[\sqrt{\frac{\epsilon_2-m_2}{\epsilon_2+m_2}}+\left(\frac{\epsilon_2-m_2}{\epsilon_2+m_2}
\right)^{3/2}+\left(\frac{\epsilon_2-m_2}{\epsilon_2+m_2}\right)^{5/2}+...\right],
\end{equation}
which allows you to save, if necessary, the symmetry of the particles.

There are two groups of relativistic corrections to the production cross sections \eqref{eq:sech1} 
connected with relative quark momenta p and q.
First group is connected with different relativistic factors in the production amplitude \eqref{eq:amp2}
containing relative momenta of heavy quarks ${\bf p}$ and ${\bf q}$. They are presented in cross
section \eqref{eq:sech1} in terms of parameters $\omega_{nk}$. 
It is important to emphasize that all these parameters can be calculated numerically within 
the quark model itself. In this sense, the proposed approach is self-consistent.
Despite the convergence of the integrals \eqref{eq:intnk} determining the relativistic parameters, when calculating them, 
we introduce an additional cutoff for relativistic momenta near the mass of c-quark $m_c$. 
The reason for this is that in the field 
of such relativistic momenta, the wave function is already very small and is not defined accurately enough, 
since its calculation uses the non-relativistic Shr\"odinger equation. Relativistic corrections
of second group are determined by bound state wave functions of pseudoscalar and vector $B_c$ mesons 
$\Psi^0_{B_{\bar bc}}({\bf q})$. For their calculation with the account of relativistic corrections
we use corresponding QCD generalization of the Breit Hamiltonian in the center-of-mass 
reference frame \cite{repko1,pot1,pot11,pot3,capstick,isgur}:
\begin{equation}
\label{eq:breit}
H=H_0+\Delta U_1+\Delta U_2,~~~H_0=\sqrt{{\bf
p}^2+m_1^2}+\sqrt{{\bf p}^2+m_2^2}-\frac{4\tilde\alpha_s}{3r}+(Ar+B),
\end{equation}
\begin{equation}
\label{eq:breit1}
\Delta U_1(r)=-\frac{\alpha_s^2}{3\pi r}\left[2\beta_0\ln(\mu
r)+a_1+2\gamma_E\beta_0
\right],~~a_1=\frac{31}{3}-\frac{10}{9}n_f,~~\beta_0=11-\frac{2}{3}n_f,
\end{equation}
\begin{equation}
\label{eq:breit2}
\Delta U_2(r)=-\frac{2\alpha_s}{3m_1m_2r}\left[{\bf p}^2+\frac{{\bf
r}({\bf r}{\bf p}){\bf p}}{r^2}\right]+\frac{2\pi
\alpha_s}{3}\left(\frac{1}{m_1^2}+\frac{1}{m_2^2}\right)\delta({\bf r})+
\frac{4\alpha_s}{3r^3}\left(\frac{1}{2m_1^2}+\frac{1}{m_1m_2}\right)({\bf S}_1{\bf L})+
\end{equation}
\begin{displaymath}
+\frac{4\alpha_s}{3r^3}\left(\frac{1}{2m_2^2}+\frac{1}{m_1m_2}\right)({\bf S}_2{\bf L})
+\frac{32\pi\alpha_s}{9m_1m_2}({\bf S}_1{\bf S}_2)\delta({\bf r})+
\frac{4\alpha_s}{m_1m_2r^3}\left[\frac{({\bf S}_1{\bf r})({\bf S}_2{\bf r})}{r^2}-
\frac{1}{3}({\bf S}_1{\bf S}_2)\right]-
\end{displaymath}
\begin{displaymath}
-\frac{\alpha_s^2(m_1+m_2)}{m_1m_2r^2}\left[1-\frac{4m_1m_2}{9(m_1+m_2)^2}\right],
\end{displaymath}
where ${\bf L}=[{\bf r}\times{\bf p}]$, ${\bf S}_1$, ${\bf S}_2$ are spins of heavy quarks,
$n_f$ is the number of flavors, $\gamma_E\approx 0.577216$ is
the Euler constant. To improve an agreement of theoretical hyperfine splittings in $(\bar bc)$ mesons
with experimental data and other calculations in quark models we add to the Breit potential \eqref{eq:breit2}
the spin confining potential obtained in \cite{repko1,repko2,repko22,repko222}:
\begin{equation}
\Delta V^{hfs}_{conf}(r)=
f_V\frac{A}{8r}\left\{\frac{1}{m_1^2}+\frac{1}{m_2^2}+\frac{16}{3m_1m_2}({\bf S}_1{\bf S}_2)+
\frac{4}{3m_1m_2}\left[3({\bf S}_1 {\bf r}) ({\bf S}_2 {\bf r})-({\bf S}_1 {\bf S}_2)\right]\right\},
\end{equation}
where we take the parameter $f_V=0.9$. For the dependence of the
QCD coupling constant $\tilde\alpha_s(\mu^2)$ on the renormalization point
$\mu^2$ in the pure Coulomb term in~\eqref{eq:breit} we use the three-loop result \cite{kniehl1997}
\begin{equation}
\tilde\alpha_s(\mu^2)=\frac{4\pi}{\beta_0L}-\frac{16\pi b_1\ln L}{(\beta_0 L)^2}+\frac{64\pi}{(\beta_0L)^3}
\left[b_1^2(\ln^2 L-\ln L-1)+b_2\right], \quad L=\ln(\mu^2/\Lambda^2).
\end{equation}
In other terms of the Hamiltonians~\eqref{eq:breit1} and \eqref{eq:breit2} we use
the leading order approximation for $\alpha_s$. The typical momentum transfer scale in a
quarkonium is of order of double reduced mass, so we set the renormalization scale $\mu=2m_1m_2/(m_1+m_2)$ and
$\Lambda=0.168$ GeV, which gives $\alpha_s=0.265$ for $(\bar bc)$ meson.
The coefficients $b_i$ are written explicitly in \cite{kniehl1997}.
The parameters of the linear confinement potential $A=0.18$ GeV$^2$ and $B=-0.16$ GeV were
obtained in quark models and lattice calculations \cite{rqm1,rqm11,rqm111,godfrey}.

\begin{table}[h]
\caption{Numerical values of relativistic parameters \eqref{eq:parameter}
in pair $B_c$ meson production cross section \eqref{eq:sech1}.}
\bigskip
\label{tb1}
\begin{ruledtabular}
\begin{tabular}{|c|c|c|c|c|c|c|}
$B_c$ &$n^{2S+1}L_J$ &$M_{B_{\bar b c}}$, GeV&$\Psi^0_{B_{\bar b c}}(0)$, GeV$^{3/2}$ & $\omega_{10}$ &$\omega_{01}$ &
$\omega_{\frac{1}{2}\frac{1}{2}}$   \\
meson  &     &    &    &    &     &      \\    \hline
$B_{\bar bc}$&$1^1S_0$ & 6.276 & 0.250 & 0.0489 & 0.0060  & 0.0171     \\  \hline
$B^\ast_{\bar bc}$  & $1^3S_1$  & 6.317 & 0.211 & 0.0540  & 0.0066   &  0.0188   \\  \hline
\end{tabular}
\end{ruledtabular}
\end{table}

\begin{table}[h]
\caption{Nonrelativistic and relativistic production cross sections of $B_c$ mesons.}
\bigskip
\label{tb2}
\begin{ruledtabular}
\begin{tabular}{|c|c|c|c|}
Final state & Center-of-mass&Nonrelativistic  cross   & Relativistic cross \\
$B_{\bar bc}+B_{b\bar c}$ &  energy s  &   section $\sigma_{nr} $ &  section $\sigma_r $ \\  \hline
$B^+_{\bar bc}+B^-_{b\bar c}$ & 22.0 GeV & $0.03\times 10^{-3}$ fb    &  $0.02\times 10^{-3}$ fb  \\  \hline
$B^{\ast +}_{\bar bc}+B^{-}_{b\bar c}$ & 22.0 GeV& $0.18\times 10^{-3}$ fb     & $0.11\times 10^{-3}$ fb\\  \hline
$B^{\ast +}_{\bar bc}+B^{\ast -}_{b\bar c}$ & 22.0 GeV& $1.43\times 10^{-3}$ fb & $0.46\times 10^{-3}$  fb\\  \hline
\end{tabular}
\end{ruledtabular}
\end{table}

\begin{figure}[t!]
\centering
\includegraphics[width=7.5 cm]{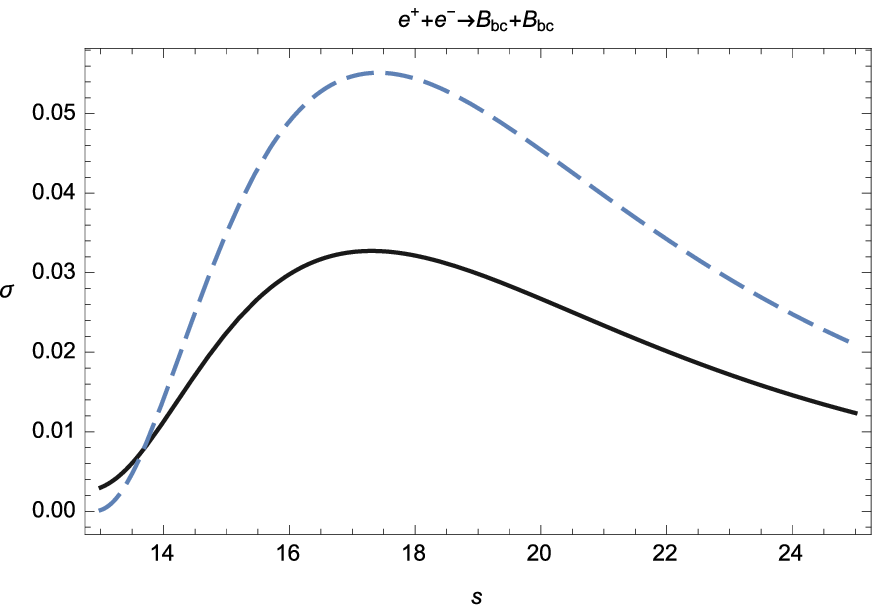}
\includegraphics[width=7.5 cm]{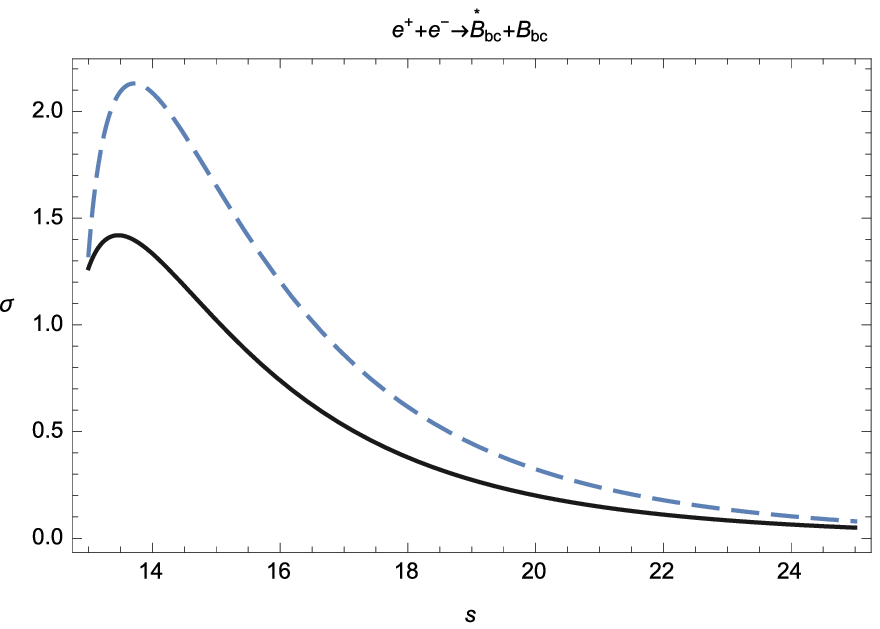}
\includegraphics[width=7.5 cm]{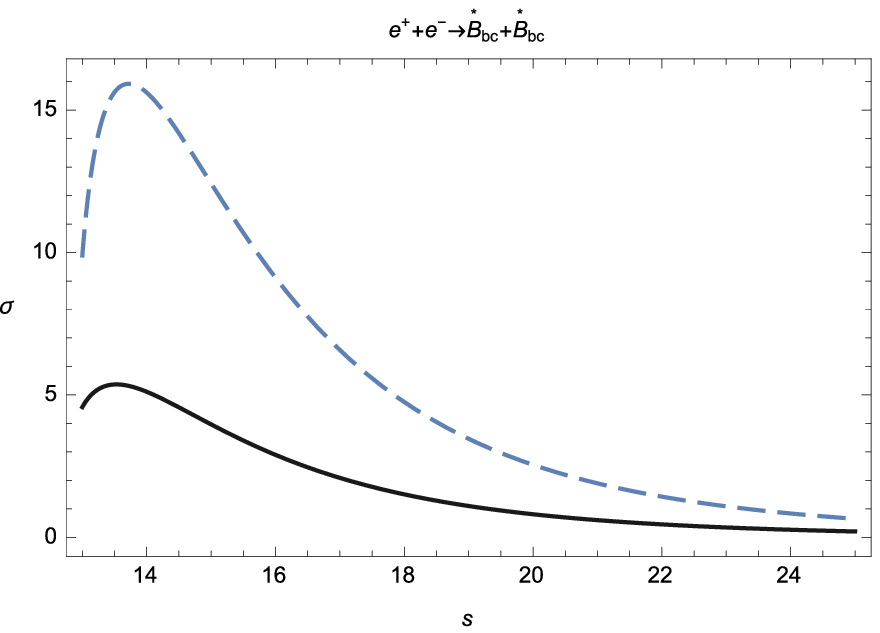}
\caption{The cross section $\sigma$ in $10^{-3}fb$ of $e^+e^-$ annihilation into a pair
of pseudoscalar and vector $B_c$ meson states as a function of the center-of-mass energy
$s$ in GeV (solid line). The dashed line shows nonrelativistic result without
bound state and relativistic corrections.}
\label{fig:fig2}
\end{figure}

To calculate relativistic corrections of second group to the pseudoscalar and vector $B_c$-meson wave functions
$\Psi^0_{B_{\bar bc}}({\bf p})$ we take the Breit potential~\eqref{eq:breit} and
construct the effective potential model as in~\cite{em2010,lucha} by means of
the rationalization of kinetic energy operator.
Numerical values of relativistic parameters entering the cross section \eqref{eq:sech1}
are calculated on the basis of \eqref{eq:intnk} and by means of
numerical solution of the Schr\"odinger equation \cite{ls,sch} and presented 
in Table~\ref{tb1}. On the basis of this model we calculate a number of
observed quantities such as the masses of charmonium, bottomonium
and $B_c$ mesons and compare the results with existing experimental data and other theoretical predictions.
The obtained results are in good agreement with them (the accuracy amounts to about one percent).
For example, in the case of low lying $(b\bar c)$ mesons we obtain $M(1^1S_0)=6.276$ GeV and $M(1^3S_1)=6.317$ GeV
(compare with experimental value $M(1^1S_0)=6.2749$ GeV \cite{pdg} and $M(1^3S_1)=6.332$ GeV \cite{efg2003}).
The similar situation occurs for charmonium states \cite{em2010}. Our estimates
of the charmonium massess agree with experimental data with more than a per cent accuracy \cite{em2010,pdg}.
Our nonrelativistic values of $B_c$ meson wave functions at the origin differs on 20 per cent from the 
values presented
in \cite{rqm1,godfrey,gklt}. 
Taking the values of parameters of the $B_c$ mesons, we
calculate the production cross sections as functions of center-of-mass energy $s$.
The plots of total production cross sections of pair $B_c$ mesons are presented in Fig.~\ref{fig:fig2}.
In Table~\ref{tb2} we give numerical values of total production cross sections at 
$s=22~GeV$ and compare them with nonrelativistic result in our quark model.
The effect of relativistic corrections to the bound state wave functions (the Breit potential)
plays a key role in total decreasing of the production cross sections as compared with nonrelativistic
results. The decreasing of nonrelativistic values of cross sections in our model with regard to relativistic 
corrections ranges from 40 to 70 percent.

\section{Numerical results and conclusion}

The studies in many previous papers 
\cite{bodwin1,bodwin2,bodwin3,chao2,chao3,bll1,bll2,em2006,ji,jia,gong,akl1,efgm2009} 
have convincingly shown that in such reactions 
as the production of bound states of heavy quarks, relativistic corrections should be taken into 
account in order to obtain more accurate values of the observed production cross sections.
Therefore, in this work we investigate not only another mechanism for the production 
of a pair of $B_c$-mesons, but also carried out an account of relativistic effects in the production 
cross section.
We develop our formalism, which was used in previous work \cite{apm2016}, in case of two-photon processes. 
In this case, the general structure of the relativistic amplitudes of the pair $B_c$ meson production 
remains the same, but the vertex functions change.
The important role of relativistic effects in exclusive processes of $B_c$ pair production 
from two-photon electron-positron annihilation is confirmed in this case. 
It is useful to say once again that when constructing the relativistic production amplitude
\eqref{eq:amp1} we keep two types of relativistic corrections which act in different directions.
The first type corrections can be 
qualified as relativistic corrections to the amplitude connected with the relative quark momenta ${\bf p}$ and ${\bf q}$.
The corrections of second type appear from the perturbative and nonperturbative
treatment of the quark-antiquark interaction operator which leads to the modification of the quark bound state wave
functions $\Psi^0_{B_{\bar bc}}({\bf p})$ as compared with nonrelativistic case.
We also systematically account for the bound state corrections working with masses of $B_c$ mesons.
The calculated masses of $B_c$ mesons agree well with previous theoretical results and experimental data
\cite{pdg,rqm1,godfrey,gklt}. Note that the quark model, which we have used in the calculations is based
on quantum chromodynamics and has certain characteristics of universality.

Total cross sections for the exclusive pair production of pseudoscalar and vector $B_c$ mesons
in $e^+e^-$ annihilation can be obtained from \eqref{eq:sech1}
after angular integration in the form:
\begin{equation}
\label{eq:sech1t}
\sigma_{{\cal PP}}=\frac{32\pi^3\alpha^4 q_c^2q_b^2M_{{\cal P}}^2|{\bf P}|M^4}{45r_1^6r_2^6s^{15}}
|\Psi^0_{{\cal P}}(0)|^4\Bigl[48\tilde s^4-24\tilde s^6+3\tilde s^8-
\end{equation}
\begin{displaymath}
12\omega_{\frac{1}{2}\frac{1}{2}} (16\tilde s^4-8\tilde s^6+\tilde s^8)+
4\omega_{01}(320\tilde s^2-144\tilde s^4+12\tilde s^6+\tilde s^8)-
12\omega_{10}(64\tilde s^2-80\tilde s^4+28\tilde s^6-3\tilde s^8)+
\end{displaymath}
\begin{displaymath}
\tilde B_P \tilde s^4\Bigl(-\frac{75 r_2 \tilde s^4}{2 r_1}-\frac{75 r_1 \tilde s^4}{2 r_2}-\frac{3 \tilde s^4}{r_1
r_2}+\frac{309 r_2 \tilde s^2}{2 r_1}+\frac{309 r_1 \tilde s^2}{2 r_2}+
\frac{12 \tilde s^2}{r_1
r_2}-\frac{288 r_2}{r_1}-\frac{288 r_1}{r_2}+75 \tilde s^4-309 
\tilde s^2+576\Bigr)
\Bigr].
\end{displaymath}

\begin{equation}
\label{eq:sech2t}
\sigma_{{\cal PV}}=\frac{32\pi^3\alpha^4 q_c^2q_b^2M_{{\cal V}}M_{{\cal P}}|{\bf P}|M^4}{9r_1^6r_2^6s^{15}}
|\Psi^0_{{\cal P}}(0)|^2|\Psi^0_{{\cal V}}(0)|^2
\Bigl[
\tilde s^6 ( 24 - 96 r_1 + 96 r_1^2 )+ 
\end{equation}
\begin{displaymath}
\tilde s^8 ( 3 - 12r_1 + 12r_1^2 )
+ \omega_{\frac{1}{2}\frac{1}{2}} (448\tilde s^4-32\tilde s^6 -8 \tilde s^8)
+ \omega_{01} ( \tilde s^4 ( 704 - 4288 r_1 + 4352 r_1^2 )+
\end{displaymath}
\begin{displaymath}
\tilde s^6 ( 272 - 864 r_1 + 768 r_1^2 )+ \tilde s^8 ( 8 + 4r_1 - 8r_1^2 ) )
+ \omega_{10} ( \tilde s^4 (- 256 - 320 r_1 + 1280 r_1^2 )+
\end{displaymath}
\begin{displaymath}
\tilde s^6 ( 304 - 1184 r_1 + 1152 r_1^2 )+ \tilde s^8( 36 - 116 r_1 + 88 r_1^2 ))-
12(\tilde B_P+\tilde B_V)(r_1-r_2)^2\tilde s^4(\tilde s^2+6)
\Bigr],
\end{displaymath}
\begin{equation}
\label{eq:sech3t}
\sigma_{VV}=\frac{32\pi^3\alpha^4 q_c^2q_b^2M_{{\cal V}}^2|{\bf P}|M^4}{45r_1^6r_2^6s^{15}}
|\Psi^0_{{\cal V}}(0)|^4\Bigl[144 \tilde s^4+ 168\tilde s^6+39 \tilde s^8+
\end{equation}
\begin{displaymath}
\omega_{\frac{1}{2}\frac{1}{2}} (-3008 \tilde s^4-1376 \tilde s^6+52 \tilde s^8)
+ \omega_{01} (3840 \tilde s^2+ \tilde s^4 ( 6272 - 320 r_1^{-1} )+ \tilde s^6 ( 2064 - 160 r_1^{-1} )
+ 52 \tilde s^8)+ 
\end{displaymath}
\begin{displaymath}
\omega_{10} (-2304 \tilde s^2+ \tilde s^4 ( 640 - 320 r_2^{-1} )+ \tilde s^6 ( 2192 - 160 r_2^{-1} )
+ 468 \tilde s^8)+
12\tilde B_V(12-41\tilde s^2-13\tilde s^4)\tilde s^4
\Bigr],
\end{displaymath}
where $\tilde s=s/M=s/(m_1+m_2)$.

The plots of total cross sections for the production of
two pseudoscalar, pseudoscalar and vector and two vector $B_c$ mesons
as functions of center-of-mass energy $s$ are presented in Fig.~\ref{fig:fig2}.
A comparison of numerical results of the two-photon mechanism of electron-positron annihilation 
with the one-photon mechanism shows that the two-photon contribution to the total cross sections 
for the production of a pair is suppressed by a small factor $10^{-3}$, which is associated with 
the ratio of interaction constants $\alpha^2/\alpha_s^2$
and qualitatively corresponds to the results for $D^+ D^-$ \cite{chao2010}.
There are no factors that could lead to an 
increase in the cross sections for two-photon annihilation compared with one-photon annihilation. 
At the production of $B_c$ mesons, there is only a recombination mechanism shown in Fig.~\ref{fig:fig1}, 
while the fragmentation mechanism of the production is absent in contrast to the production 
of a pair of charmonia. 
The developed technique will be applied to the $e^+ e^-\to B_cB_c e^+ e^-$ process, which is not 
suppressed by the propagators, and for which it can be expected that its contribution will be 
comparable to the contribution of one-photon annihilation, as shown for pair production
$D^+D^-$ in \cite{berezhnoy}.
Note also that we are discussing a comparison of the relativistic 
and nonrelativistic cross sections in the tail section of the plots in Fig.~\ref{fig:fig2}, 
in which the $M^2/s^2$ corrections of higher order are strongly suppressed.
Accounting for the interference effects from the amplitudes of one-photon and two-photon annihilation 
gives terms of order $\alpha^2\alpha_s^2\frac{\alpha}{\alpha_s}$ in the differential cross section, 
which essentially has the same order 
of smallness as the radiative corrections of order $O(\alpha_s)$ to the amplitudes of one-photon 
annihilation. But since such terms are proportional to odd powers of $z=\cos\theta$  in the differential 
cross section, their contribution to the total cross section is zero.

From the results presented in Fig.~\ref{fig:fig2}, it follows that
an account of relativistic and bound state corrections decreases the
values of nonrelativistic production cross sections. 
We would like to emphasize that the term nonrelativistic limit means that $M_{B_{\bar bc}}=m_1+m_2$,
and the wave function of the bound state of quarks was determined by solving the
Schr\"odinger equation with a purely nonrelativistic Hamiltonian.
In the resulting expressions for relativistic cross sections, there are various relativistic 
factors that affect the change in the original non-relativistic section in different ways.
The greatest reduction in cross sections is given by the value of bound state wave function at the origin.
Relativistic effects in the production amplitude, which are determined by the parameters $\omega_{10}$, $\omega_{01}$, 
$\omega_{\frac{1}{2}\frac{1}{2}}$ give an increase in the cross sections by several tens of percent.
It is useful to recall that effects of order $O(\alpha_s)$ also increase the values of the cross sections.

As is well known \cite{brambilla2011,bbl}, calculations similar to those given in this paper contain 
a number of theoretical uncertainties.
Our calculation of production cross sections of a $B_c$ meson pair via two-photon annihilation mechanism
is based on relativistic quark model, which can be considered as a microscopic theory of
quark-gluon and photon interactions. The quark model allows you to perform the calculation
of observables and different parameters describing the formation of bound states of heavy quarks
can be found within its framework. This property of the quark model represents its obvious advantage.
In this work we take into account corrections of the second order in relative momenta p and q.
The used calculation method can be extended to include fourth-order corrections in accordance with p and q. 
The arising new parameters can also be estimated within the framework of the quark model itself. 
But in this calculation, these corrections are included in the theoretical calculation error, 
which we define at 30 percent.
Another important theoretical calculation uncertainty is associated with the determination 
of the wave functions of the bound states of quarks in the region of relativistic momenta $m_c$.
We estimate the total error of this type at $5\%$. 
Then the corresponding error in determining the cross sections \eqref{eq:sech1} for the production 
of a pair should not exceed $20\%$. 
In our opinion, this rather approximate estimate is consistent with calculations of the mass spectrum 
of heavy quarkonia, in which a large error in determining the wave functions in the field of relativistic 
momenta will give a discrepancy with the observed masses of more than one percent.
An important part of the total theoretical error is connected with radiative corrections of order $\alpha_s$, 
which are not considered in this work. 
It can be said that the use of the Breit Hamiltonian in the calculation of the wave functions 
leads only to a partial account of corrections of this type.
We assume that such radiative corrections may give a 20 percent change in the production cross sections.
The total maximum theoretical error can be estimated in $40\%$.
To get it, we add the above estimates in quadrature.

\acknowledgments
The work is supported by Russian Foundation for Basic Research (grant No. 18-32-00023) (F.A.M. and O.S.S.).
A.V. Berezhnoy acknowledges the support from "Basis" Foundation (grant No. 17-12-244-1).

\appendix

\section{The coefficient functions $f^{i,rel}$, $f^{bind}$ and $f^{LO}$ entering in
the $B_c$ meson production cross section (16)}

\vspace{5mm}

{\underline {$e^++e^-\to B_{\bar bc}^++B_{b\bar c}^{-}$}.

\begin{equation}
f^{LO} =16\tilde s^4z^2(1-z^2 )-8\tilde s^6 z^2(1-z^2 )+\tilde s^8 z^2 (1-z^2 ),
\end{equation}
\begin{equation}   
f^{1,rel}=-64\tilde s^4 z^2 (1-z^2 )+
32\tilde s^6 z^2(1 -z^2 )-4\tilde s^8z^2(1-z^2 ),
\end{equation}
\begin{equation}   
f^{2,rel}=\frac{1280}{3}\tilde s^2z^2(1-z^2)
-192\tilde s^4z^2(1-z^2)+6\tilde s^6z^2 (1-z^2)
+\frac{4}{3}\tilde s^8z^2 (1-z^2),
\end{equation}
\begin{equation}   
f^{3,rel}=-256\tilde s^2z^2(1-z^2)+320\tilde s^4z^2(1-z^2)
- 112\tilde s^6z^2(1-z^2)+12\tilde s^8z^2 (1-z^2),
\end{equation}
\begin{equation}   
f^{bind}=\frac{\tilde s^4}{r_1r_2}[z^2 (-96 (r_1 - r_2)^2 + 4 \tilde s^2 + 44 (r_1 - r_2)^2 \tilde s^2 - \tilde s^4 - 
5 (r_1 - r_2)^2 \tilde s^4 + 
\end{equation}
\begin{displaymath}    
96 (r_1 - r_2)^2 z^2 - 4 \tilde s^2 z^2 - 
39 (r_1 - r_2)^2 \tilde s^2 z^2 + \tilde s^4 z^2)].
\end{displaymath}

\vspace{5mm}

{\underline {$e^++e^-\to B^{\ast +}_{\bar bc}+B^-_{b\bar c}$}.

\begin{equation}
f^{LO} =( 1 - 4r_1 + 4r_1^2 )[4\tilde s^6+\tilde s^8+4\tilde s^6z^2-\tilde s^8z^2],
\end{equation}
\begin{equation}
f^{1,rel}=
\frac{256}{3}\tilde s^4-\frac{8}{3}\tilde s^8
+\frac{128}{3}\tilde s^4z^2-\frac{64}{3}\tilde s^6z^2+\frac{8}{3}\tilde s^8z^2,
\end{equation}
\begin{equation}
f^{2,rel}=\frac{1}{3} \tilde s^4( 256 - 1984 r_1 + 2176 r_1^2 )+
\frac{1}{3}\tilde s^6 ( 224 - 800 r_1 + 704 r_1^2 )+
\frac{4}{3}\tilde s^8 ( 2 + r_1 -2 r_1^2 )+
\end{equation}
\begin{displaymath}    
\frac{1}{3}\tilde s^4 ( 640 - 2624 r_1 + 2176 r_1^2 )z^2+
\frac{32}{3}\tilde s^6 (  -4 + 21 r_1 - 18 r_1^2 )z^2
-\frac{4}{3}\tilde s^8 (2 + r_1 - r_1^2 )z^2,
\end{displaymath}
\begin{equation}    
f^{3,rel}=\frac{64}{3}\tilde s^4(-1 - 5 r_1 + 10 r_1^2 )
+\frac{1}{3}\tilde s^6( 128 - 608 r_1 + 704 r_1^2 )
+\frac{4}{3}\tilde s^8 (9 -29 r_1 + 22 r_1^2 )+
\end{equation}
\begin{displaymath}    
\frac{320}{3}\tilde s^4(  -1 + r_1 + 2 r_1^2 )z^2
+\frac{32}{3}\tilde s^6(7 -17 r_1 + 6 r_1^2 )z^2
+\frac{4}{3}\tilde s^8(- 9 + 29 r_1 - 22 r_1^2 )z^2,
\end{displaymath}
\begin{equation}   
f^{bind}=[-2(3 +3 z^2 + \tilde s^2 (1 - z^2))](r_1-r_2)^2 \tilde s^4.
\end{equation}

\vspace{5mm}

{\underline {$e^++e^-\to B^{\ast +}_{\bar bc}+B^{\ast -}_{b\bar c}$}.

\begin{equation}
f^{LO} =48\tilde s^4(1 - z^2 )z^2+ 8\tilde s^6 (1 - 2z^2 + 3z^4 )
+ \tilde s^8 ( 2 + z^2 - 3z^4 ),
\end{equation}
\begin{equation} 
f^{1,rel}= \frac{64}{3}\tilde s^4 (-10 + 13 z^2 - 3 z^4 )
+\frac{32}{3}\tilde s^6(- 4 - 7 z^2 + 3z^4 )+\frac{4}{3}\tilde s^8( 2 +  z^2 -3z^4 ),
\end{equation}
\begin{equation}   
f^{2,rel}=1280\tilde s^2 z^2(1 -z^2 )
+\frac{1}{3}\tilde s ^4 (896 - 128 r_1^{-1} z^2 - 1216 z^2 + 1728z^4 )+
\end{equation}
\begin{displaymath}    
+\frac{16}{3}\tilde s^6 ( 18 - 2 r_1^{-1} + 2 r_1^{-1} z^2 + 3z^2 -9 z^4 )
+\frac{4}{3}\tilde s^8 ( 2 + z^2 - 3z^4 ),
\end{displaymath}
\begin{equation} 
f^{3,rel}=-768\tilde s^2z^2(1 -z^2 )
+\frac{1}{3}\tilde s^4 (- 128 - 128 r_2^{-1} z^2 + 2368 z^2 -2880z^4 )+
\end{equation}
\begin{displaymath}    
+\frac{1}{3}\tilde s^6( 288 - 32 r_2^{-1} + 32 r_2^{-1} z^2 - 592 z^2 + 1008 z^4)
+12\tilde s^8 ( 2 + z^2 -3 z^4 ),
\end{displaymath}
\begin{equation}   
f^{bind}=[(-16 \tilde s^2 - 
8 \tilde s^4 + (-96 + 40 \tilde s^2 - 4 \tilde s^4) z^2 + (192 - 96 \tilde s^2 + 12 \tilde s^4) z^4)]\tilde s^4.
\end{equation}

\end{document}